# Why Are Cuprates the Only High-Temperature Superconductors?

J. C. Phillips

Dept. of Physics and Astronomy,

Rutgers University, Piscataway, N. J., 08854-8019

## ABSTRACT

The hierarchical mean-field theory of elastic networks, originally developed by Maxwell to discuss the stability of scaffolds, and recently applied to atomic networks by Phillips and Thorpe, explains the phase diagrams and remarkable superconductive properties of cuprates as the result of giant electron-phonon interactions in a *marginally unstable mechanical network*. The overall cuprate networks are fragile (floppy), as shown quantitatively (with an accuracy ~ 1%), and *without adjustable parameters*, by comparison with stabilities of generically similar network glasses, and are stabilized by percolative backbones composed of isostatic $CuO_2$ planes.

PACS indices: 74.72.-h  74.62.-c  74.25.Kc  74.25.Dw

## 1. Introduction

To answer the titled question one should first answer the question of what is the mechanism responsible for superconductivity in the cuprates. The conventional answer is electron-phonon interactions, just as in other superconductors, but there is an abundance of parameterized models based on other mechanisms. Because of the structural complexity of the ceramic cuprates, the relevance of e-p interactions (or any of the unconventional alternatives) is not easily established. In elemental metallic superconductors, convincing evidence was obtained from isotope shifts of $T_c$. Several observables in the cuprates show large isotope shifts [1], but the isotope shifts of $T_c$ often decrease and become quite small at optimal doping. This is understandable if one assumes that optimization involves many factors, and that at optimal doping these other factors can (and should) compensate isotope shifts [2]. Similarly, fine structure in I-V tunneling characteristics confirms the Eliashberg theory of e-p interactions and the gap equation in elemental metallic superconductors, and similar fine structure has been observed in cuprates, but only rarely [3]. It is now understood that the reason for this is that for most tunnel junctions fine structure is not resolved because of gap inhomogeneities [4]. Once again the complexity of the cuprates has obstructed the establishment of e-p interactions as the mechanism for superconductivity. The gap inhomogeneities give rise to a multitude of anomalies (ten or twenty or perhaps even more), and isolated mechanisms can be constructed to explain one or two of these at a time, but no global picture has appeared so far.

There is a way to resolve dilemmas of this kind: because complexity and strong disorder obstruct the conventional path, one adopts a *global* approach that *utilizes* those factors. One steps back from detailed microscopic pictures of "strong Coulomb interactions" [severely truncated and lavishly parameterized Hubbard toy σ models, etc.] on an atomic scale, which are nothing more than statements of the obvious, that binding in transition metal oxides is a complex mixture of covalent and ionic bonding, to examine the elastic stability of the host network. Such an examination may provide us with insight into many of the key questions of the cuprates, the most important of which is why they are dopable, and why upon doping they are the only oxides to exhibit HTSC, instead of some other anomalous property (such as colossal magnetoresistance, CMR). (Of course, this kind of *lattice* approach does not address the anomalous *electronic* properties that appear in both the normal and superconductive states upon doping; these are a separate issue that the author's filamentary model has addressed in great detail; see below.)

By focusing on the elastic properties of cuprate hosts one avoids the very large task of explaining the anomalies generated by the electronic dopant network, as well as the many very severe limitations of mean field many-body electronic models. The phenomenology of the cuprates provides us with many reasons to suppose that such a program, properly executed, can be successful. The many electronic anomalies apparently are made possible not only by the large number of atoms in the unit cell, but by the even larger number of unit cells in a "checkerboard" nanodomain of average diameter 3nm (ten unit cells) [4]. These large numbers mean that there are generally many ways that one can claim to have given a qualitatively correct description of one or two aspects of HTSC (such as one or two features of their phase diagrams) using methods that would apply equally well, and give quite similar results, for almost any host lattice, for instance, a manganite with CMR instead of a cuprate with HTSC. The lack of chemical *specificity* in such schematic models, reflected by generous helpings of adjustable parameters, greatly reduces their significance.

A general characteristic of all perovskite and pseudoperovskite oxides is that they are subject to a very wide variety of lattice instabilities. Lattice instabilities are commonly observed in metallic alloys, and are often the factor that limits $T_c$. However, there are many instabilities in oxides, yet most of them are not dopable, and although the manganites can be rendered metallic by alloying, they cannot be made superconductive. Thus it is apparent that the phrase "lattice instabilities" is itself much too general (much like "strong Coulomb interactions"), but how are we to make it more *specific*, when the number of such possible instabilities is very large? The answer to this question is that we need a generic theory that describes the tendency towards all (or nearly all) possible instabilities, subject only to the restriction that the instability leads to a space-filling network structure. Mathematically this may seem to be impossible, but physically nature has already given us the answer. Atomic networks with a large variety of instabilities and local topologies are achieved in good glass formers, in other words, network glasses, which are space-filling when quenched from the melt. The mean-field theory of elastic networks, originally developed by Maxwell to discuss the stability of scaffolds, and recently applied to glassy atomic networks [5] by Phillips and Thorpe, then explains the remarkable superconductive properties of cuprates as the result of giant electron-phonon interactions in a *marginally unstable mechanical network*. The overall network is stabilized by isostatic CuO2 planes. [In hydrodynamics isostatic means in hydrodynamic equilibrium. Here it means a special kind of mechanical equilibrium defined below.]

This model explains the bulge in the phase diagrams of ab basal areas of CuO2 planes of LSCO and Hg cuprates associated with the intermediate superconductive phase [6]. The general shape of this bulge is quadrilateral, with a gradual rise above background with doping as Tc increases to its maximum value, followed by a sharp drop in the second spinodal region (Fig. 1). This bulge may indeed reflect the extra electronic effects of carriers on basal areas of CuO2 planes. These Fermi-energy electronic effects are large just because of the isostatic (marginally stable) elastic character of the CuO2 planes. The gradual increase at lower doping reflects the increasing electronic pressure on the isostatic backbone of the percolative electronic paths responsible for high temperature superconductivity [7]. Because these paths thread through dopants placed *between* isostatic $CuO_2$ planes, locally near dopants the c axis contracts, leading to the observed expansion (bulge) of basal areas that keeps the average unit cell volume nearly constant. There are similar correlations of $T_c$ with buckling of the cubic lattice constants of the transition metal nitrides (such as NbN), but there the buckling can have either sign [8]. These superconductive correlations are unrelated to the presence of magnetic nanodomains [5], which represent an incidental secondary phase.

The general theory of space-filling elastic networks with hierarchical forces [6] has identified **three** phases: the two obvious ones, which are underconstrained (floppy) and overconstrained (rigid). These are the two phases that one would expect to find with short-range forces only. However, because elastic forces are long-range, there is a third phase, which is exactly constrained (so not floppy), but with no excess or redundant constraints. This third phase is only a critical point in mean-field models of the phase diagram, but it becomes a separate and distinct phase, spanning a range of compositions, when allowance is made for nanoscale inhomogeneities: then it is identified with an isostatic backbone, which is locally neither under- nor over-constrained [6]. The three phases, including the isostatic backbones, have been identified in extremely accurate and complete studies of network glasses, and in later sections we will draw on these for *quantitative*, *parameter-free* estimates of the elastic properties of cuprates. The third phase is the intermediate phase that is responsible for the remarkable filamentary electronic properties of cuprates, both in the superconductive and normal states [7].

Broadly speaking, the stability and long-range connectivity of such electronic filaments is greatly enhanced when they are embedded in the isostatic intermediate atomic phase.

These very strong topological parallels between the intermediate phases in network glasses and in cuprates provide, in the author's view, much more justification for the following discussion than one could ever hope to achieve from any calculation based on toy model or even realistic Hamiltonians, because even the latter have so far not proven to be capable of identifying and describing the intermediate phase in network glasses, which are in many respects much more easily understood than are the strongly disordered cuprates. Moreover, the alert reader will notice in the following discussion something quite striking: there are *no adjustable parameters*. How, he may be thinking, is such a thing possible? How is it possible that a global theory can be constructed that is of quantitative value without so much as mentioning very large Coulomb repulsions, etc.? The answer is that the constraint theory discussed below is hierarchical, and it easily focuses on relative energies in any given energy range (here the range of phonon energies, which is only a few % of the much larger correlation energies). The important point, which is well known to organic chemists in the context of the π bonding (Hueckel) theory of hydrocarbons, is that for energies of order $T_c$ or T, the strongest σ interactions are not relevant in discussing chemical trends, it is the marginal π interactions that are critical. However, even with hierarchical ordering a method must still be devised to make certain that all interactions are properly counted; this is done successfully here *without adjustable parameters* because many of the microscopic details are subsumed in topological data carried over from the properties of other known (and much simpler!) network structures. This is the way that Pauling discussed heats of formation in molecules and crystals, and it is here applied to the properties of elastic networks. It turns out (and the author himself has found this very surprising) that the apparently extremely delicate issues of network lattice instabilities are handled at least as successfully here as Pauling was in his treatment of heats of formation. It may well be that the reason for this success is that the latter are more sensitive to the non-transferable many body core polarization effects that are described by adjustable parameters in the Hubbard σ approach. Those parameters are not needed here, as bonds are never really broken, but are merely bent ($\pi$ interactions). The Hubbard $\sigma$ terms could be important in discussing the phase transition between the antiferromagnetic insulator and the metallic phase, but this transition is of no concern here, as antiferromagnetism is a common phenomenon in transition metal oxides, and hence it is much too general to be related to the specificity of HTSC in the cuprates.

## 2. CuO$_2$ planes

The most popular explanation for the uniqueness of the cuprates is based on the observation that all contain CuO$_2$ planes. Moreover, small buckling of these planes (on a scale of 1%) significantly reduces [9] T$_c$ (see Fig. 2). This has led most analysts to conclude that the very large electronic interactions responsible for HTSC are concentrated in these planes. Here we argue, however, that this conclusion is simplistic. First, the cuprates that contain *only* CuO$_2$ planes (the La$_{2-x}$B$_x$CuO$_4$ family, B = Ca, Sr, Ba) have substantially *lower* maximum T$_c$'s (similar to MgB$_2$) than those that contain other metallic planes, such as secondary planar arrays of CuO$_{1-x}$ chains, or BiO or HgO planes. This very basic fact is inexplicable by models that ascribe HTSC to electronic interactions in CuO$_2$ planes. However, all those other metallic planes are mechanically much softer than the CuO$_2$ planes, which are mechanically rigid. (The planar lattice constants of all the cuprates are very similar, which shows that CuO$_2$ planes fix the planar lattice constants because they are elastically the strongest planar elements in these multilayer structures. Softening of secondary planes relative to CuO$_2$ planes is discussed below.) Generally speaking, chemical trends in metallic and intermetallic superconductors have shown that increased screening of ion-ion repulsions by electron-ion interactions has two effects: increases in T$_c$, and reductions (softening) of observed vibrational frequencies. Thus if the very large interactions responsible for HTSC are in fact conventional electron-phonon interactions, then these interactions will be weakest in the metallic CuO$_2$ planes, and much larger in the softer and less mechanically stable arrays of metallic CuO$_{1-x}$ chains, or BiO or HgO planes, accounting for the higher T$_c$'s of compounds containing the latter. The CuO$_2$ planar buckling is indeed one of the keys to HTSC, but not for electronic reasons; quite the contrary, for mechanical reasons, as we shall see.

High resolution scanning tunneling microscope studies [10] of $CuO_2$ planar spectra of metastable $CuO_2$ terraces on BSCCO show a 60 meV gap, with a rather broad peak in the density of states, very similar to the broad 60 meV pseudogaps (possibly Jahn-Teller gaps) observed in STM on BSCCO BiO natural cleavage planes [4,11]. They see *no* evidence in the $CuO_2$ planes for a narrow superconductive gap peak near 40 meV. Instead, there appears to be a 10 meV insulating region around $E_F$, that may be the result of surface buckling of the metastable terraces. STM studies of the $CuO_{1-x}$ surface chains of YBCO revealed [12] a rich superconductive subgap structure probably associated with vacancies in the O chains. Near edge XAFS on detwinned $Y_{1-x}Ca_xBa_2Cu_3O_{7-y}$ single crystals, where the Ca has supposedly introduced holes only into the $CuO_2$ planes, led to the conclusion that superconductivity arises only when there are holes in the planes *and* the chains and the interplanar apical oxygens as well [13]. Thus experimental data lead to the conclusion that electronically the rigid $CuO_2$ planes function primarily as electrical connectors between softer metallic elements where the strong electron-phonon interactions actually occur.

## 3. Defects

Cuprates become metallic only upon doping, and like many other strongly ionic doped semiconductors, are subject to compensation of dopants by defects. In particular, the defects can easily condense to relieve interlayer misfit stress by forming insulating nanodomain walls, thereby destroying metallic planar character. This can happen very easily in doped anionic insulators, especially in oxides, where the anionic interactions are large, but also where the oxygen anions are exceptionally mobile. One can then suppose that the function of the $CuO_2$ planes is to suppress such nanoscale catastrophes in marginally unstable layer structures. The problem for theory, then, is to show why the $CuO_2$ planes are ideally suited to performing this function, while buckling occurs in almost all other layered oxides of the very numerous pseudoperovskite families of, for example, dozens of other $A_2BO_4$ compounds [14]. In particular, what is the mechanical mechanism that makes the $CuO_2$ planes so stable against defect formation?

This is not a question that is easily answered in sufficient generality by conventional analytic methods based on concrete multi-parameter interatomic force fields fitted to observed vibrational spectra [3], as these exhibit many low-frequency ($< 100$ cm$^{-1}$) modes. However, it can be resolved by combining more abstract methods involving (1) general concepts of chemical bonding that accurately explain chemical trends in partially ionic – covalent bonds, and (2) topological techniques that determine the generic stability of partially covalent bonding networks for glasses and defective crystals. (1) General trends in chemical bonding of $A^NB^{8-N}$ compounds show that Cu, alone among the monovalent alkali and noble metals, is sufficiently covalent to have well-defined covalent bonds, although the Cu halides are marginally stable because they lie very close to the covalent-ionic transition and accordingly have very small shear elastic constants [15]. (2) Topological techniques describe the stiffness transition from floppy to rigid very accurately in network glasses [16]. The most surprising aspect of these techniques is that they are able to describe the properties of a *space-filling* disordered network entirely by algebraic methods (linear equations in a mean-field approximation). The method involves comparing degrees of freedom with covalent bond-bending and bond-stretching constraints. It presumes a very high degree of disorder (as is found in network glasses), and may be modified in crystalline applications to stiff, well-ordered materials like mono- and di-silicates [17]. STM experiments [4,10,11] have revealed very strong planar electronic disorder in the cuprates, and thus it is plausible that topological techniques that have been successful for network glasses can describe trends in electronic defect chemistry in the cuprates; at present there are no reasonable alternatives that can treat strong disorder.

## 4. Counting Network Elastic Constraints: Isostatic Rigidity

The *space-filling algebra* is implemented as follows. We count covalent bond-bending and bond-stretching constraints. The bond stretching constraints $\alpha$ are easy: they are just $C_s = m/2$ for each atom (no double counting). What about the bond-bending angular constraints $\beta$? How many pairs of bonds are there? Of course, $C_{ap} = m(m-1)/2$, but how many of these are linearly independent? Suppose we place a **local polar axis** at each atom, whose orientation is specified by the d coefficients of the d = 3 Cartesian unit vectors. The bond-bending energies $\sim \beta \mathbf{R}_{ij} \bullet \mathbf{R}_{jk}$ will be determined when the (d - 1) polar angles ($\theta$, $\phi$,...) are known for each of the m bonds, or (d-1)m constraints. However, the orientation of the polar angle is arbitrary, so there are really only $C_{ia} = (d-1)m - d$ linearly independent angular constraints. (Check: for d = 3 and m = 2, there is one bond-bending constraint. This holds for $m \geq d - 1$. For $m < d - 1$, $C_{ap}$ is correct.)

At this point we need to step back and look at the idea of the **local polar axis** quite carefully. This may seem to be a natural idea, but it is not the only possibility. For instance, we could have defined a common polar axis for the entire plane, and oriented each local polar axis relative to the sample polar axis: this gives a reduction of only 2 bending constraints per atom, not 3. How do we know which way is correct? In the network glass case there is no crystalline order, so local polar axes are certainly the correct way to describe the glassy structure. In the cuprate case there is always some buckling[2], and it is clear from the strong disorder observed by STM that the buckling varies significantly from site to site. At this point all one can say is that the assumption of strong local buckling (*independent* rocking of $CuO_6$ octahedra) gives much better agreement with experiment than alternative assumptions. This assumption actually contains two parts, buckling of Cu - O - Cu bond angles, and buckling of O - Cu - O bond angles. Note that it is the O atoms that buckle out of plane to first order, but the effect of buckling on the O - Cu - O bond angles is only second order (the octahedra rock, but are not distorted). We are therefore justified in counting the Cu bending constraints as intact, while the O bending constraints are broken. All the nearest neighbor stretching constraints are intact, of course. Broken oxygen bending constraints were first identified [18] in a glassy context for $g-SiO_2$; they are a characteristic feature of oxides, and may be the single most important reason why there is so much disorder even in crystalline oxides.

The number of intact planar valence force field constraints per atom is thus determined by m = 4 for Cu, which gives 7 = [2 (stretching) + (8 − 3) (bending)] Cu constraints, and 2 = [2 (stretching) + 0 (bending)] constraints for the two O atoms. Altogether this is 9 constraints per $CuO_2$, or 3 per atom. Thus the number of constraints per atom is equal to the number of degrees of freedom (d = 3), which is the mean-field condition for an ***isostatic (rigid but unstressed)*** elastic network. When this condition is satisfied, strain energy is minimized in the mean-field approximation by a defect-free network. In fact, interlayer misfit (which has been neglected) will still generate some defects, but weaker interlayer effects are complex and cannot be treated by available methods. They would involve selective displacements of apical oxygen atoms associated with defects or dopants. Such displacements have been observed [19,20] by XAFS near Sr and Ni impurities in double-doped crystals of $La_{2-x}Sr_xCu_{1-y}Ni_yO_4$) at x = 0.15 for several values of y spanning the metal-insulator transition. One of the striking results of the studies is that the $NiO_6$ octahedra are contracted along the c axis by approximately 0.32 Angstrom relative to $CuO_6$ octahedra, while the in-plane distances of $NiO_6$ and $CuO_6$ octahedra are the same within 0.01 Angstrom ($CuO_2$ planar rigidity). The c axis distortions show a break in slope across the metal-insulator transition, which is dramatic evidence for the isostatically (that is to say, marginally) rigid character of $CuO_2$ planes. This is confirmed by studies of anomalous zone-boundary $CuO_2$ planar longitudinal optic phonons that show that these phonons appear in $La_{2-x}Sr_xCuO_4$ (LSCO) just at x = 0.06 (the metal-insulator transition) [21]. Similar LO phonon anomalies are found in all the HTSC [22,23], and they leave no doubt that it is the electron-phonon interaction that is responsible for HTSC.

The reader who is not interested in technical details, and does not want to invest much time in studying the foregoing discussion at length, can still understand its success. The important point to appreciate is that reasoning of this kind has a firm mathematical foundation [6] that asserts principles that are generic, that is, that apply equally well to *any* strongly disordered, space-filling network. The ideas have been tested in applications to more than ten binary and ternary chalcogenide and oxide network glass alloys [24,25]. The local topologies of these alloys are varied and complex, mixtures of corner- and edge-sharing tetrahedra, pyramids, fragments of linear and spiral chains, etc. Regardless of these varied local topologies, the topological counting procedure is accurate and completely reliable: it has always been successful in describing the phase diagrams of these disordered networks *without adjustable parameters*. Moreover, it is the only method that does account for these phase diagrams. Given that cuprates are themselves strongly disordered, there is good reason to expect success here as well, even though the presence of longer range forces in these weakly metallic compounds implies even greater complications in terms of packing and space filling.

## 5. Alternative Constraint Counting

One way to bring out the physical content of constraint models is to consider alternatives. Here the natural alternative to a mechanical model that focuses specifically on the local rigidity of isostatic $CuO_2$ planes is a mean field model of the entire lattice. In mean field theory, with all bending constraints intact, the isostatic average coordination number is 2.40. There are hundreds of compounds with perovskite or distorted perovskite structures, but the most common ones, which occur in nature or can be prepared synthetically, are exemplified by $(Ca,Sr,Ba)TiO_3$ and $PbZrO_3$, which have the chemical formula $A^{II}B^{IV}O_3$. The average valence of these compounds is 2.40, and so they are isostatic (strain-free) on the average. This provides a natural explanation for why it is so easy to grow large crystals of these materials, and why they are so popular in a wide variety of applications. The reader may object that average valence is the same as average coordination number only in the case where all bonds are single covalent bonds, but this objection is readily met, as we shall now see.

HTSC cuprate structures typically consist of cuprate planes (weakly metallic, but largely covalent) alternating with ionic planes (LaO, SrO, …) and other metallic-covalent planes (cuprate chains, BiO, HgO). It seems unlikely that a basic mechanism could be contained in such a hodgepodge; that is why attention has focused on the common factor, $CuO_2$ planes. Moreover, counting constraints for such mixtures of all kinds of chemical bonds appears to be impossible. However, constraints have been counted for the window glass mixture ($SiO_2$, $Na_2O$, and CaO) that contains both ionic and covalent bonds [24]. The count was very successful, as it predicted the composition of this extremely common, very important, and indeed unique ternary commodity to within 1%, *using no adjustable parameters*.

The principles involved in counting covalent constraints require mainly that one distinguish between broken and intact bending constraints of single bonds. Ionic constraints are different, because a monovalent element like Na is often 6- or 8-fold coordinated. Pauling solved this problem by introducing the concept of resonating bonds, and he demonstrated that this way of counting produces reasonably good consistency between molecular and crystalline heats of formation (bonding energies). Thus monovalent Na is regarded as having a coordination number (or number of equivalent single bonds) of one, regardless of its actual coordination number. Thus if the coordination number is 6, one says that there is 1/6 of a single bond to each ligand: the single bond "resonates" from ligand to ligand. (This is the *correct usage* of Pauling's term "resonating valence bond".)

If we supplement the covalent counting rules with Pauling's concept, one simply replaces the average coordination number with the average valence, and can count constraints for a mixed ionic – covalent network as before, remembering only to correct for broken single bond bending constraints. For example, for $La_2CuO_4$ the average valence should be counted as $16/7 = 2.28 < 2.40$; average valences for the other HTSC cuprates are similar but lower (most of the elements are divalent: Cu, O alkaline earths, with usually one trivalent element), spanning the range down to 2.00 (Hg cuprates, all elements divalent). For a mean field calculation this result is quite reasonable – it says that the overall lattice is floppy relative to the isostatic perovskites $A^{II}B^{IV}O_3$, and that it has been stabilized by $CuO_2$ planes, which act as isostatic (rigid but unstressed) backbones. Boolchand has identified a number of such isostatic backbones in the reversibility windows of singly bonded chalcogenide glass alloys [25]. The windows are found to lie in the range of coordination numbers $2.27 – 2.52$. Because so many alloy systems have been studied, this range appears to be well established with an accuracy $\pm 0.03$ ($\sim 1\%$).

Thus the *upper end* of the floppy HTSC cuprate range [2.00, 2.28] *just touches the lower end* of the Boolchand isostatic mean field chalcogenide glass range [2.27, 2.52], and the $CuO_2$ planar backbones are just barely necessary and sufficient to provide marginal isostatic stability. This places the cuprates at the limit of glassy elastic stability, which is just where one would expect to find the highest temperature superconductors.

The question of dopability involves both elasticity and electronic filaments, and so lies somewhat outside the scope of this paper, but the following points can be made. In semiconductors normally electrically active dopants are substitutional (simple acceptors and donors, with valence one less or one more than the atom they replace). However, replacement of Cu in $CuO_2$ planes by Ni (nominally an acceptor) or Zn (nominally a donor) does not alter the apparent charge density, but it rapidly depresses $T_c$. This behavior seems very mysterious from the point of view of mean field theories that ascribe all the electrical properties of the cuprates to the $CuO_2$ planes. In fact, electrically active dopants are always located *outside* the $CuO_2$ planes. There are two reasons for this: an atomic one (the planes outside the $CuO_2$ planes are floppy, and little strain accompanies placing dopants in these soft planes), and an electronic one (the dopants acts as bridging elements that enable filaments to thread between metallic domains in metallic planes). Finally, HTSC itself is made possible because at these dopants electron-phonon interactions are very large just because the dopants are embedded in an underconstrained, anomalously soft environment.

## 6. Buckling and Threading

There are some subtleties of Fig. 1 that are quite interesting. Looking only at the experimental basal areas $A(x)$ one sees a corner at compositions between $x_0$ and $x_2$. It is tempting to identify this corner with a rounded two-dimensional saddle point logarithmic singularity in an effective density of states for filamentary states. However, strictly speaking one should subtract the background and look at $\delta A(x)$. The background, as extrapolated here following Rohler, also contains a corner at $x_0'$, this time between $x_1$ and and $x_0$. The difference $\delta A(x)$ resembles a mesa with two shoulders at $x_0'$ and $x_0$, which is typical of three-dimensional densities of states with two reversed square root singularity saddle points adjacent to the band center. Note also that $\delta A(x)$ strongly resembles $T_c(x)$ in LSCO with its two spinodal immiscibility domes [7]. These similarities suggest that the quadrilateral shape of $\delta A(x)$ reflects the three-dimensional nature of threading filaments, even though $A(x)$ itself is a two-dimensional quantity, a kind of magical step-up in dimensionality. One must remark that large amounts of hitherto hidden information can be uncovered in apparently simple X ray data, as carefully analyzed, for example, by Rohler [5].

### 7. Chemical Trends in Electron-Phonon Interactions

Constraint counting also provides an amazingly simple and apparently very accurate assessment of the *relative* strength of electron-phonon interactions at dopants. In the cuprates the dopants are usually oxygen, or Sr in LSCO, and the dopant states may involve $CuO_2$ as well, as these dominate the band structure near $E_F$. These dopants are divalent and lie at the bottom of the marginally soft cuprate range [2.00, 2.28], and are well below the Boolchand glass range [2.27, 2.52]. One can contrast this with the situation in the manganites, where unmistakable evidence for a filamentary "ghost" metal in $La_{2-2x}Sr_{1+2x}Mn_2O_7$ has been obtained by photoemission [26]. The existence of the filamentary network (the "anti-Jahn-Teller effect") [27] is consistent with the *low* average valence of 2.33. The structure of this compound consists of rigid $MnO_2$ layers separated by (La,Sr)O double layers containing the natural cleavage plane. Most of the states near $E_F$ are centered on $MnO_2$ clusters, with average valence 2.67. These clusters are certainly rigid, as their average valence lies well *above* the Boolchand glass range [2.27, 2.52]. The resulting electron-phonon interaction will be too weak to produce superconductivity, even in metallic filaments, as transverse to the filaments there are large unscreened repulsive Coulomb interactions.

It would appear that similar reasoning applies to the superconductive perovskites ($T_c \sim$ 40K) of the $(Ba,K)(Pb,Bi)O_3$ family, where the average valence is also 2.40. However, this apparent exception actually supports, rather than disproves, the constraint approach.

The point is that in the column IV series C, Si, Ge, Sn, Pb, with increasing principle quantum number n the bond-bending forces that are strong for small n are progressively weakened, leading to the borderline structure of white Sn, and broken bond-bending constraints in the Pb row. In the absence of the latter, the number of constraints per atom in the $(Ba,K)(Pb,Bi)O_3$ family is reduced, and the average valence corresponding to this reduced number of constraints is only 2.00. In fact, it seems likely that this is a borderline case, but the source of stronger e-p interactions in this family compared to the manganites is clear: it is the weakening of bond-bending constraints at the heavy (Pb,Bi) sites. The weakening of bond-bending constraints in the heavy Bi or Hg secondary metallic planes contributes similarly to the enhancements of $T_c$ in BSCCO and the Hg cuprates. Finally, in $YBCO_7$ the reduction of planar coordination from 4 in the $CuO_2$ planes to only 2 in the CuO chains greatly softens the latter, resulting in the higher $T_c$ in YBCO compared to LSCO.

## 8. Conclusions

Because electron-phonon interactions are so strong in perovskites and especially in the layered pseudoperovskites, it is reasonable to relate the dopability and the marginally metallic character of the cuprates to the ideally isostatic character of $CuO_2$ planes. In other words, the intermediate phase [7] that is responsible for HTSC in the cuprates is an isostatically rigid phase so far as only the $CuO_2$ planes are concerned. The cuprates are also floppy (only marginally stable) in the context of a mean field approximation for the lattices as a whole (all planes), much like true perovskites, which are noted for their numerous and complex ferroelastic and ferroelectric instabilities. This gives rise to strong e-p interactions at divalent dopants. By contrast, in the manganites the $MnO_2$ clusters are too rigid, relative to the Boolchand range, to produce e-p interactions large enough to overcome the large Coulomb repulsions characteristic of ionic crystals. Thus the manganites do not exhibit HTSC, but they do exhibit CMR, and they are also spatially inhomogenous, like all other perovskites.

One of the noteworthy features of the present model is that it readily explains the dramatic scalar lattice constant correlation [5] shown in Fig. 1, which is also a considerable improvement over the possibly more natural angular bending correlations [9] shown in Fig. 2. Those readers who feel that these results would be better justified through more elaborate calculations may consider the following points. No method based on adjustable parameters can explain chemical trends, especially the uniqueness of HTSC in the cuprates. Moreover, before attempting to treat the cuprates, true "first principles" quantum calculations should obviously focus on the much easier example of the NbN family, where 5% lattice bulges (4d transition metals) and dips (3d transition metals) were correlated with superconductive $T_c$ 's [8]. This correlation was made more than 30 years ago, and the capabilities of "first principles" quantum calculations have made great strides since then. However, to the author's knowledge no one has even attempted to explain the lattice buckling. (It may well be that electronic structure experts found the sign reversal between the 4d and 3d cases too daunting, especially in the light of the fact that even today most electronic structure calculations of lattice constants are not accurate to 1%. Moreover, this sign reversal indicates failure of rigid band or mean field models at some level, which at present is unknown.) At the same time, this model represents a microscopic realization of some of the intuitive factors that guided Bednorz and Mueller in making their historic discovery [28], a discovery not emulated by theorists with parameterized models. Readers who are interested in the connection between stress in the host lattice and the electronic properties of the dopant-based filamentary network embedded in it should study the recent 70-page, 55-figure review [29], which also discusses the dramatic nanodomain structure observed by STM [4,11,12]. This structure is self-organized, reflecting the enhanced stability and long-range connectivity of such electronic filaments when they are embedded in the isostatic intermediate atomic phase. The easiest way to see this is to compare correlation lengths (~ 10 nm) obtained by finite size scaling analysis [30] of the specific heat and penetration depth using a two-fluid (liquid He) model with those observed directly [4,11,12] or inferred from critical current (breakdown of self-organization) data [31], both of which give 3 nm. They should also take note of the new ARPES data on the isotope dependence of the phonon kink [32], as well as recent theoretical studies showing enhanced e-p interactions in localized states [33]. The inadequacy of models based on adjustable parameters is well known, but it has been amusingly reviewed recently [34]. In my opinion, there are many σ-π similarities between Dyson's failed multiparameter pseudoscalar meson theory of proton-meson scattering, and parameterized, Hubbardized models of HTSC.

I am grateful to J. D. Jorgensen for a copy of Fig. 2, and to J. Rohler for discussions.

**Figure Captions**

Fig. 1.  Sketch of the bulges (lightly shaded) in basal area of $CuO_2$ planes, as displayed, together with original data, in [5].   The sketch here refers to $(Y,Ca)Ba_2Cu_3O_x$ and $HgBa_2CaCu_2O_x$, where the magnitude of the bulge is about twice as large as in $La_{2-x}Sr_xCuO_4$, corresponding to stronger e-p interactions and higher $T_c$ in the former. Note that the intermediate phase bulge lies above the linearly extrapolated basal areas of the insulating and normal metallic phases.   Note also that because interactions are screened in the metallic phase, $x > x_2$, the slope is lower there than in the insulating

phase, $x < x_1$.  In the most studied case, $La_{2-x}Sr_xCuO_4$, $x_0 = 0.16$, $x_1 = 0.06$, and $x_2 = 0.21$.

Fig. 2.  The buckling of $CuO_2$ planes breaks the bond-bending constraint for $Cu - O - Cu$ bond angles, and causes departures from colinearity [9].  This breaking correlates well with depressions in $T_c$.  Note that the compound Hg 1212 has the smallest distortion and the largest $T_c$.  Considering the large differences in crystal structures, this is an excellent correlation; it foreshadowed the results of later neutron studies of phonon spectra, made possible by growth of large single crystals [21-23].

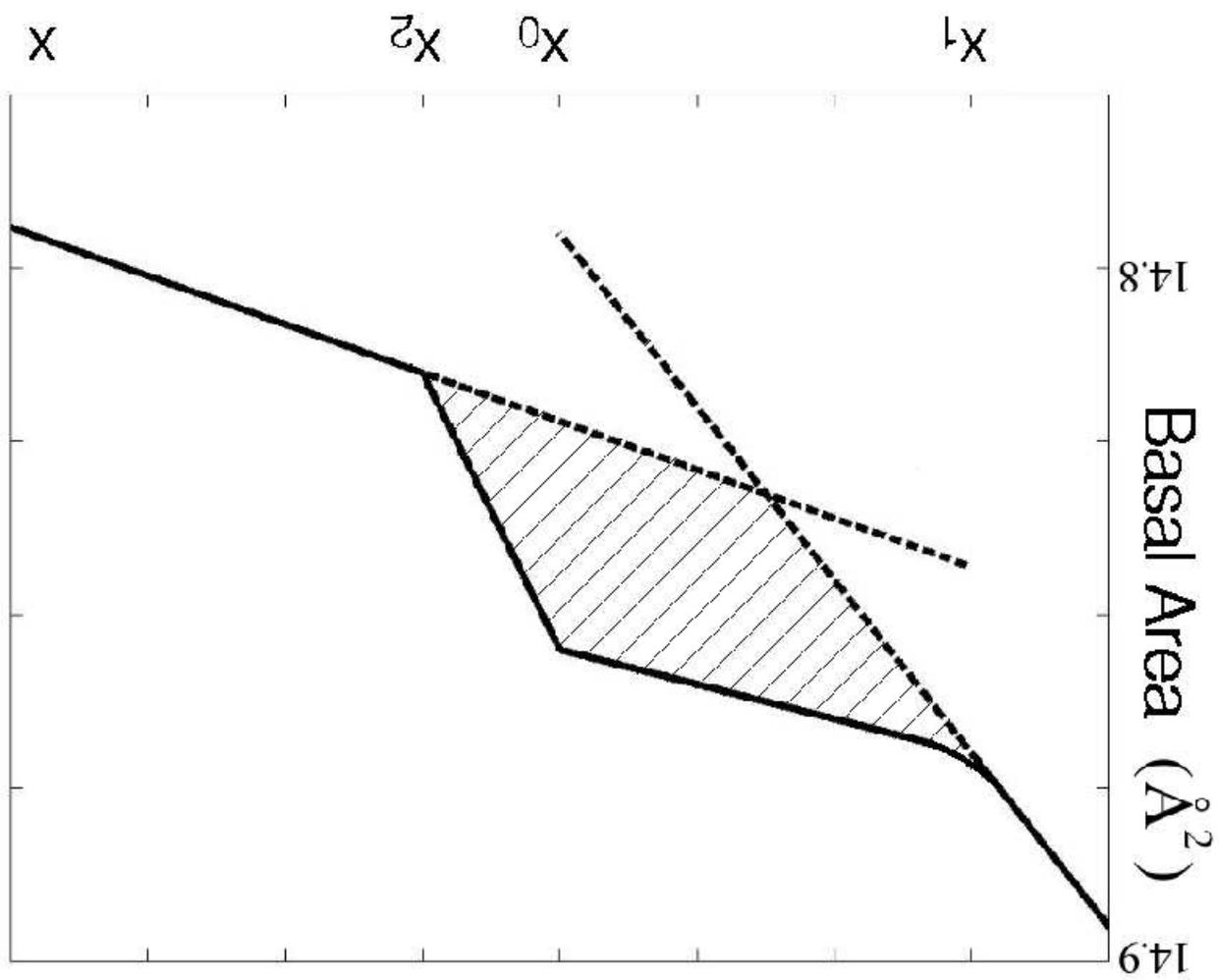

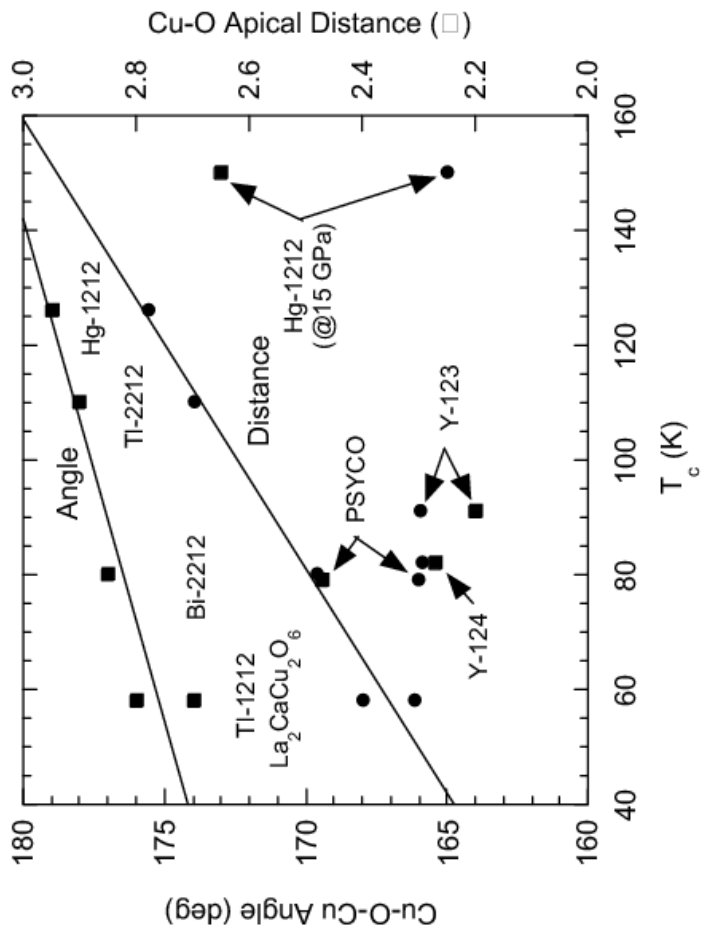